\documentclass[twocolumn,showpacs,preprintnumbers,amsmath,amssymb]{revtex4}
\usepackage{graphicx}
\usepackage{dcolumn}
\usepackage{bm}

\begin{document}
\preprint{}

\title{Fundamental thermal fluctuations in microspheres}

\author{M.L. Gorodetsky}
\author{I.S. Grudinin}
\affiliation{Moscow State University, Faculty of Physics, Moscow, Russia}
\email{gorm@hbar.phys.msu.ru}

\begin{abstract}
We present theoretical analysis and the results of measurements of
fundamental thermorefractive fluctuations in microspheres.
Experimentally measured noise spectra are consistent with the theoretical
model.
\end{abstract}



\maketitle

\section{\label{sect:intro} Introduction}

The development of modern technology in many fields leads to further miniaturization
of components. This makes it necessary to take into account certain fundamental
physical limitations. An example of such limitations are thermodynamical fluctuations
of temperature in a small volume. These fluctuations are transformed to wideband noise
in the output channel due to the temperature dependence of device parameters. The same
limitations also frequently appear in experimental physics in macroscopic
high--precision measurements. As it was recently shown, fundamental fluctuations with
the same origin limit the sensitivity of gravitational wave antenna (international
project LIGO\cite{LIGO}), where thermal expansion, thermal dependence of refractive
index and Young's modulus give rise to different types of noise
\cite{99a1BrGoVy,00a1BrGoVy,Cagnoli}. Thermorefractive fluctuations lead to phase
noise in long fibers\cite{Glenn, Wanser}, which were used to observe the effect for
the first time \cite{WanserM}.

Microspheres \cite{89a1BrGoIl} are relatively novel type of optical resonators
uniquely combining small size (from tens to thousands of micrometers) and high
quality--factor up to $Q\simeq 10^{10}$ for the so--called ``whispering--gallery
modes''\cite{96a1GoIlSa} (WGMs). The small size of the effective volume, occupied by
the e.m. field of the mode, makes for low thresholds of nonlinear effects such as
bistability and oscillatory instability \cite{92a1GoIl}, which are preconditioned by
Kerr's and thermal effects. The small volume makes it possible for such resonators to
be used as a tool for detection and measurement of the thermorefractive noise
\cite{01a1Go}. Apart from its importance in the LIGO project, the measurement of
thermorefractive noise can serve as an innovative experimental examination of the
theory of microscopic fluctuations of temperature. Thermorefractive noise should also
be taken into account in possible applications of whispering--gallery modes resonators
such as diode laser stabilization.

\section{Thermorefractive noise in microresonators}
One can understand the effect from the well known
thermodynamical equation for the variance of temperature
fluctuations $u$ in the volume $V$:
\begin{equation}
\langle u^2\rangle = \frac{\kappa T^2}{\rho C V} \label{variance},
\end{equation}
where $T$ is the temperature of the heat--bath, $\kappa$ is the Boltzmann constant,
$\rho$ is density, and $C$ is specific heat capacity. By substituting in parameters
for fused silica: $\rho=2.2\ {\mbox{g}}/{\mbox{cm}^3}$, $C=6.7\times 10^6\
\mbox{erg}/(\mbox{g}\cdot\mbox{K})$ and effective volume of field of the most
localized WGM in the microsphere of radius $R\sim50\ \mu\mbox{m}$ with
$V_{eff}\simeq10^{-9}\ \mbox{cm}^3$, we obtain the value of the standard deviation of
temperature $\sqrt{\langle u^2\rangle}\simeq 30\,\mu\mbox{K}$. These temperature
fluctuations, combined with the coefficient of thermal refraction $dn/dT=1.45\times
10^{-5}\ K^{-1}$ lead to the effect of relative eigen frequency fluctuations
$\delta\omega/\omega \sim 3\times10^{-10}$. The estimate is comparable with the
bandwidth of resonances achievable in microspheres. We do not consider thermoelastic
noise in microspheres, since the coefficient of thermal expansion $\alpha=5.5\times
10^{-7}\ \mbox{K}^{-1}$ is sufficiently smaller in fused silica than $dn/dT$. In
subsequent sections we will present more rigorous analysis aimed at finding spectral
properties of this noise, taking into account peculiar field distribution of WGM.

Variations of the refractive index $n$ in the dielectric cavity perturb its resonant
frequencies. The perturbed wave--equation has the form:
\begin{equation}
\Delta \vec E+(\epsilon^0+2n\delta n)\frac{\omega^2}{c^2} \vec
E=0,
\end{equation}
where $\vec E$ is the electric field strength in the cavity, $\epsilon^0=n^2$ is
permittivity, and $\delta n=\frac{dn}{dT} u$ is the variation of refractive index due
to fluctuations of temperature $u$. If $\vec E_0$ is the ortho--normalized field
distribution of an eigenmode of the unperturbed cavity ($\int \vec E_i \vec E_j^*d\vec
r=\delta_{ij}$) and $\omega=\omega_0+\delta\omega$ is the frequency shift then after
multiplication of this vector equation on complex conjugated vector $\vec E_0^*$ and
integration over the whole volume, neglecting the second order terms we obtain:
\begin{equation}
\frac{\delta \omega}{\omega_0}=-\frac{1}{n}\int\limits_V |\vec
E_0^2|\delta n d\vec r= -\frac{1}{n}\frac{dn}{dT}\bar u,
\label{baru}
\end{equation}
where $\bar u$ is the temperature deviation, averaged over the mode volume.

\section{Power spectral density of thermal fluctuations}
To calculate the effect of fluctuations of temperature the method of
fluctuational thermal sources $F(\vec r,t)$ may be used  \cite{99a1BrGoVy,00a1BrGoVy}:
\begin{eqnarray}
\frac{\partial u}{\partial t} -D\Delta u = F(\vec r, t), \label{TDFT}
\end{eqnarray}
where $D=\lambda^*/(\rho C)$ is thermal diffusivity and $\lambda^*$ is thermal
conductivity ($\lambda^*=1.4\times 10^5\
\mbox{erg}/(\mbox{cm}\cdot\mbox{s}\cdot\mbox{K})$, $D=9.5\times
10^{-3}\mbox{cm}^2/\mbox{s}$ for fused silica). This approach is analogous to the
Langevin approach, which uses fluctuational forces in the equations of dynamics. It
was shown before that if proper normalization of the sources is used:
\begin{eqnarray}
B^F_{rt}&=&\langle F(\vec r, t) F(\vec
r',t')\rangle\nonumber\\
&=&\frac {2\kappa T^2 D}{\rho C}\nabla^2
\delta(\vec r-\vec r')\delta(t-t'), \label{norm}
\end{eqnarray}
this approach leads to the correct results, which satisfy the Fluctuation--Dissipation theorem
(FDT). In particular, it was shown that thermoelastic noise is associated through FDT
with thermoelastic damping. It is also possible to show that thermorefractive noise is
connected through FDT with electrocaloric losses \cite{electrocaloric}.

Thermodynamical fluctuations of temperature, averaged over the mode volume,
may be calculated as:
\begin{equation}
\bar u =  \int u(\vec r, t) |E_0(\vec r)|^2 d\vec r,
\end{equation}
where $\int|\vec E_0|^2d\vec r=1$ and
\begin{equation}
u(\vec r, t)=\int
\frac{F(\vec\beta,\Omega)}{D\beta^2+i\Omega}e^{i\Omega
t+i\vec\beta\vec r}\frac{d\Omega \,d\vec\beta}{(2\pi)^4},
\label{solution}
\end{equation}
is the general solution of (\ref{TDFT}). Spectral correlations of fluctuational
forces satisfy the following condition:
\begin{eqnarray}
B^F_{\beta\Omega}&=&\langle F(\vec\beta',\Omega')F^*(\vec\beta,\Omega)\rangle \nonumber \\
&=&{(2\pi)^4}\frac{2kT^2D}{\rho C}\beta^2
\delta(\vec\beta-\vec\beta')\delta(\Omega-\Omega')\label{correl1}
\label{speccorrel}.
\end{eqnarray}
We may now calculate the following averaged value
$B^u_{\tau}=\langle \bar u(t)\bar u(t+\tau)\rangle$ (correlation
function of temperature fluctuations averaged over the mode
volume) and so, from Wiener--Hinchin, theorem the one--sided
(hence additional factor 2) power spectral density $S_{\bar
u}(\Omega)$ of fluctuations of temperature:
\begin{eqnarray}
&&S_{\bar u}(\Omega)=\frac{4 k T^2 D}{\rho C}\nonumber\\
&\times& \int\!\int\!\int\frac{\beta^2
|E_0(\vec r)|^2 |E_0(\vec r')|^2}{D^2\beta^4+\Omega^2}  e^{i\vec\beta (\vec r - \vec r')} d\vec r d\vec r' \frac{d\vec \beta}{(2\pi)^3}\nonumber\\
&=&\frac{4 k T^2 D}{\rho C}
\int\frac{\beta^2|G(\vec\beta)|^2}{D^2\beta^4+\Omega^2}\frac{d\vec\beta}{(2\pi)^3}\label{useful},
\end{eqnarray}
where
\begin{equation} G(\vec\beta)=\int|\vec E_0|^2
e^{-i\vec\beta\vec r} d\vec r
\end{equation}
is normalized spatial spectrum of the energy distribution in the resonator. To verify
this useful expression we may integrate it over all frequencies:
\begin{eqnarray}
\langle u^2\rangle=\int\limits_0^\infty S_{\bar
u}(\Omega)\frac{d\Omega}{2\pi} =\frac{k T^2}{\rho C}\int
|G(\vec\beta)|^2\frac{d\vec\beta}{(2\pi)^3}.
\end{eqnarray}
Comparing the final expression with (\ref{variance}) we deduce that
\begin{equation}
V_{eff}^{-1}=\int |G(\vec\beta)|^2\frac{d\vec\beta}{(2\pi)^3} =
\int |\vec E(\vec r)|^4\ {d\vec r}.
\end{equation}
The same general expression for the effective volume of the mode
in microresonator appears in the analysis of nonlinearity
\cite{92a1GoIl} and scattering \cite{00a1GoIlPr} in microspheres.

It is important to note that the above expressions were obtained
ignoring boundary conditions. If the field, is concentrated near thermally isolated surface,
 as in the case of microsphere, (\ref{speccorrel}) should be modified, substituting in
(\ref{correl1})
$[\delta(\beta_\bot-\beta_\bot^\prime)+\delta(\beta_\bot+\beta_\bot^\prime)]\,
\delta(\vec \beta_\Vert-\vec \beta'_\Vert)$ instead of
$\delta(\vec\beta-\vec\beta')$, where $\vec\beta_\bot$ is the
component of wave--vector of fluctuations normal to the surface
and $\vec \beta_\Vert$ are components, parallel to it.

In the analysis above, the medium was considered infinite (the volume of field
localization is significantly smaller than the size of a device). Discrete spectrum of
thermal waves should be considered for more accurate calculations, especially at very
low frequencies:
\begin{equation}
F(\vec r,t)=\int\sum F_\nu(\Omega)\Phi_\nu(\vec r) e^{i\Omega t}
\frac{d\Omega}{2\pi}.
\end{equation}
\begin{eqnarray}
B^F_{\nu\Omega}&=&\langle F_\nu(\Omega)F^*_{\nu\prime}(\Omega')\rangle\nonumber\\ &=&
\int\!\int\!\int\!\int\! B^F_{rt}\Phi_{\nu\prime}(\vec r')\Phi^*_\nu(\vec r)
e^{i(\Omega' t'-\Omega t)} d\vec r dt d\vec r' dt'\nonumber\\ &=&2\pi \frac{2\kappa
T^2 D}{\rho C} \beta^2_\nu \delta (\Omega-\Omega') \delta(\nu,\nu').
\end{eqnarray}
Analogously to  (\ref{solution})
\begin{equation}
u(\vec r, t)=\int \sum_\nu
\frac{F_\nu(\Omega)}{D\beta_\nu^2+i\Omega}\Phi_\nu(\vec r)
e^{i\Omega t} \frac{d\Omega}{2\pi},
\end{equation}
and
\begin{eqnarray}
\bar u(t)= \int \sum_\nu
\frac{F_\nu(\Omega)G_\nu}{D\beta_\nu^2+i\Omega} e^{i\Omega t}
\frac{d\Omega}{2\pi},
\end{eqnarray}
where $G_\nu$ are the coefficients of field's intensity decomposition into normal
thermal waves $\Phi_\nu(\vec r)$ of the finite medium.

Now, as before, we may calculate the correlation function of
relative frequency fluctuations and power spectral density:
\begin{eqnarray}
S_{\bar u}(\Omega)= \frac{4 k T^2 D}{\rho C}
\sum_\nu\frac{\beta_\nu^2|G_\nu|^2}{D^2\beta_\nu^4+\Omega^2}\label{useful2}.
\end{eqnarray}
Applying the obtained expressions to the microsphere leads to the following expression
for the power spectral density  of relative frequency fluctuations:
\begin{eqnarray}
S_{\delta\omega/\omega}(\Omega)&=&S_{\bar
u}\left(\frac{d\,n}{d\,T}\right)^2=\frac {\kappa
T^2\sqrt{\ell}}{\pi^{3/2} n^2R^2\sqrt{\lambda^*\rho
C\Omega}}\left(\frac{d\,n}{d\,T}\right)^2\nonumber\\
&\times&\frac{1}{\sqrt{1-b^2/d^2}}\frac{1}{(1+(\Omega\tau_b)^{3/4})^2},
\label{theory}
\end{eqnarray}
where $\ell,m,b,d,\tau_b$ are parameters, determined by the mode in microsphere (see
{\bf Appendix A}). This expression was experimentally verified in this work.

It is essential to note that in a finite body $S_{\bar u}(\Omega)$, contrary to
(\ref{theory}), remains limited at zero frequency. However, instead of the sum
(\ref{sums}), we used the former expression (\ref{theory}), obtained from the
continuous spectrum, to evaluate the results of the measurements. There were important
reasons for doing this. First of all, too many approximations were used when obtaining
the sum. These are crude approximations especially for lower frequencies. However, it
is shown in the Appendix and confirmed by numerical calculations that the sum
(\ref{sums}) and integral (\ref{ints}) lead to the same asymptotic dependence at high
frequencies. Secondly, microspheres during fabrication are formed on short fused
silica stem which conducts heat and couples microspheres to heat--bath. Thermal
exchange with the atmosphere due to convection is also not negligible. Therefore,
microspheres are not thermally isolated. Thirdly, other parasitic noise effects
dominate at low frequencies. Throughout the experiment, we only considered this
phenomenon at frequencies of $\Omega>1/\tau_R=D/R^2\sim 400 c^{-1}$ for ($R=50\mu m$).

\section{Experimental Setup}\label{sect:measurement}
The idea of measurement of thermorefractive noise in optical microspheres is quite
simple. If one tunes the measuring laser's frequency on the slope of resonance curve
of WGM, that is on the range, where amplitude of the signal being measured sharply
depends on frequency, then the trembling of resonator's eigen frequency transform into
fluctuations of intensity of output radiation. These intensity fluctuations are
recorded and further processed.
\begin{figure}[htbp]
\includegraphics{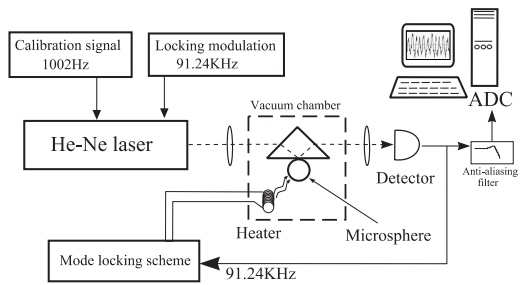}
\caption{\label{fig:epsart} Scheme of experimental setup for the
measurement of noise spectra in microspheres.}
\end{figure}
To measure thermorefractive noise we used small $2R\simeq 80\div 260\mu m$ fused
silica microspheres with quality factors of $Q\simeq 10^9$, manufactured with the
hydrogen miniburner. In order to avoid degradation of Q--factor due to adsorption of
atmospheric water \cite{96a1GoIlSa} microspheres were placed  upon fabrication to the
atmosphere of dry clean nitrogen into the special chamber where all measurements were
conducted.

WGM in microspheres were excited with the prism coupler using He--Ne laser
($\lambda=0.63\mu m$) with piezo--driven front mirror as in \cite{94a1GoIl}. Laser
frequency could be tuned within the range of approximately $0.8GHz$. The output power
of laser ($540\mu W$) was reduced to $45\mu W$ passing into the chamber. The reduction
of power was necessary to weaken thermal and Kerr nonlinearities \cite{92a1GoIl} and
was carried out with the neutral glass filters. Frequency of the laser was locked at
the slope of resonance curve. This was done with the use of mini--spiral heater,
placed nearby the resonator, which actively stabilized the temperature of the
microsphere, hence stabilizing its eigen frequency. To provide the feedback loop a
weak signal at $91.2kHz$ was used to modulate laser frequency. This frequency was
chosen to coincide with one of acoustic resonances of the front mirror to facilitate
laser frequency modulation. This signal is demodulated by the WGM and error signal was
used to correct the current which feeds the heater. When the WGM is locked to the
laser, the output of the microsphere comprises the thermorefractive amplitude noise.
The output from the resonator was registered with Thorlabs PDA500 amplified GaAsP
photodetector, with $40dB$ gain in amplitude and $f_d=45kHz$ bandwidth at this gain
setting.

The signal from detector was digitized using ADC (ICP DAS PCI--$1802L$ computer board)
with digitization speed of 333 kilosamples per second  and 12--bits per sample
resolution. Continuous sets of $6553600$ ($19.66s$) points were recorded in each
session of measurement for further processing in each session of measurement. To find
the estimate of noise frequency spectrum, the record was subdivided into many equal
time intervals. Power spectral densities were calculated using Fast Fourier Transform
(FFT) algorithm for each interval and then averaged over the entire set of intervals
within each session. 100 intervals with $65536=2^{16}$ points in each were used to
produce estimate of spectra with fine resolution of $\approx 5Hz$. To avoid the effect
of frequency aliasing, the frequency band of the signal incoming from  the detector
was limited by the Butterworth filter of the eighth order. At stop--band frequency of
$166kHz$ (the Nyquist frequency in our case) this filter attenuates the signal to
-48dB level. First order RC--filter with time constant of $\tau\simeq 2s$ was used to
get rid of the constant offset.

\subsection{Calibration of spectra}
\label{sect:calibration}

In order to obtain absolute values of spectral densities for measured spectra, we have
developed and used the method of calibration based on laser frequency modulation. In
this method, weak sinusoidal voltage of known amplitude and frequency is admixed to
piezo--actuator of laser's mirror, thereby producing weak laser's frequency
modulation. Since the measured spectra represent relative frequency fluctuation, this
additional sinusoidal modulation of laser frequency results in a narrow peak in
frequency spectrum. If the amplitude of frequency modulation is known, it is possible
to find the value of spectral density corresponding to the spike.

\begin{figure}[htbp]
\scalebox{0.5}{\includegraphics {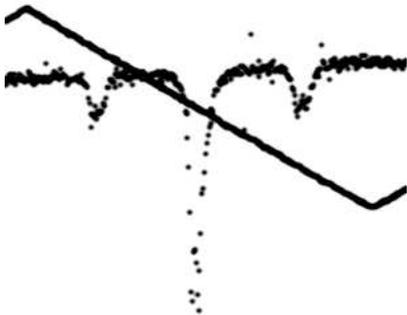}} \caption {\small  Oscillogram of
frequency marks obtained with sideband modulation technique and saw--tooth voltage at
the  piezo--actuator as used for calibration of actuator and quality--factor
measurement. The frequency of modulation is $5MHz$ and the corresponding Q--factor,
obtained from resonant curve width is $7\times10^8$} \label{fig:fmarkse2}
\end{figure}

To find the coefficient of transformation of voltage to laser frequency modulation and
to measure the quality factors of the resonators, we performed calibration of
piezo--actuator using following technique. If laser frequency is swept by the
saw--type voltage with known amplitude in the vicinity of eigen frequency of the
 microsphere and additional phase modulation with frequency higher than the
bandwidth of resonant curves is applied to the pump, one can see on the detector two
additional resonances apart from the central frequency.
 These sidebands produce
frequency scale for the calibration of piezo--actuator and for the measurement of
Q--factor from the width of resonant curves (see fig.\ref{fig:fmarkse2}).
The coefficient of piezo--actuator has been measured and is equal to $\frac{\partial
f}{\partial v}=(3,63\pm 0.1)\frac{\footnotesize{MHz}}{\footnotesize{V}}$.

If the amplitude of frequency modulation produced by
piezo--actuator is known, the amplitude of the peak it produces in
calculated power spectral densities can be used for their absolute
calibration (see {\bf Appendix B}).

To produce calibration peak we applied $5\div 15mV$ modulation at $172.9Hz$ or
$1002Hz$ (preferred frequency at later stages of experiment) to piezo--actuator of
laser mirror, which resulted in relative frequency modulation of
$P=\delta\omega/\omega=(4\div11)\times10^{-11}$. These frequencies were chosen to
match that of bins of discrete Fourier transform of 65536 points (namely
$1001.99=197\cdot F_s/65536$ and $172.93=34\cdot F_s/65536$). The higher frequency of
$1kHz$ was chosen so that the calibration peak to be separated from low frequency
setup noises in the frequency spectrum. It is important to note that the described
calibration method does not require the quality factor of the microsphere as well as
the transformation coefficients and gains in electrical tracts to be known.

After amplitude calibration is performed, additional correction is required to
compensate for frequency filters in the detector by dividing the resulting spectrum by
$1/\sqrt{1+(f/f_d)^2}$ --- the frequency response of the detector. Frequency response
of anti--aliasing filter was also compensated. To do this, it was measured and
approximated with the polynomial of 7--th order so that the compensation procedure
could be carried out by point--to--point multiplication.

\subsection{Excess noises in the measurements}
\label{sect:hindrances} In this research, the superfluous noises were represented by
the electrical interference (generally $50Hz$ and harmonics), acoustical and seismical
noises, electromagnetic foil of devices and also by inherited amplitude and frequency
noises of the laser. The frequency noise of the laser has turned out to be the  most
essential source of influence on the quantity being investigated. We have measured
this noise together with all other technical noises of the setup. The same technique
as for thermorefractive noise measurements was used with only difference of using
relatively large microspheres with diameters of $481, 570, 508, 588, 894, 619 \mu m$
and modes with large $l-m$ numbers of the order of 100 and above.  In these modes,
thermorefractive noise appears to be small enough for laser noise to dominate.
Measurements of the laser noise for all aforementioned microspheres coincided well
enough. The microspheres operated as the frequency discriminators in this case.

Laser noises calibration was carried out in the same way as it was done for
thermorefractive noise calibration --- with the use of calibrating peak. In
measurements of the technical noise spectra calibrating spike at frequency of $172Hz$
was used.

Special measures were undertaken to weaken electromagnetic hindrances in the setup. We
have applied proper signal shielding, moreover, we have shielded ADC board inside the
computer, which resulted in the setup noise decrease of more than order of magnitude.
The dynamic range of digitization system exceeded $90dB$ (taking into account
averaging over many spectra).

\subsection{Identification of WGMs and computer processing}
Upon completion of recording process, the identification of WGM indices was carried
out. The knowledge of the modes' indices allows one to calculate the effective volume,
occupied by the electromagnetic field of the mode, as well as theoretical energy
distribution within the volume and therefore the theoretical spectral density. To
perform the identification, filming with the digital camera of the speckled image of
the mode was carried out. The speckled image is called forth by scattering of mode's
e.m. field at the residual molecular inhomogeneities on the surface of the microsphere
\cite{96a1GoIlSa}.
\begin{figure}[htbp]
\scalebox{0.22}{\includegraphics{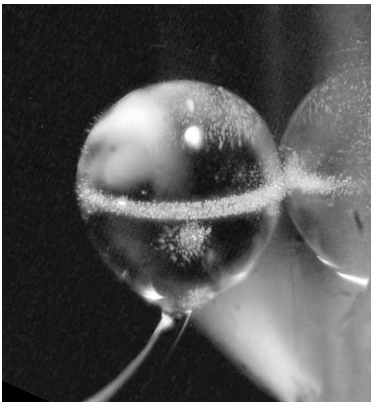}} \caption{\small The resonator $570\mu
m$ in diameter with $l-m\simeq13$ mode (magnification 88x).}
\end{figure}
The approximation formulas were derived, linking the microsphere's
radius with $\ell$ index and the angular half--width of the WGM's
belt with $\ell$ and $\ell-m$ numbers. (See {\bf Appendix C}).

Upon recording, images were magnified and processed with the made--on--purpose
computer program of graphical identification of the WGMs. The belt's width ($\ell-m$
value), simulating sphere size, slope and tilt of the belt could be controlled
interactively, so that the theoretical framework projection of the mode could be made
coinciding with the photographic projection of the real mode. This allowed one to
obtain the parameters of the mode. The error of estimate of the $\ell-m$ value is
proportional to $\sqrt{\ell-m+1/2}$ i.e. the width of the mode belt, but is strongly
determined by the quality of the pictures obtained. Modes with $\ell-m\leq3$ can
usually be precisely recognized while for $3<\ell-m<6$ the error in obtaining the
indices approaches 1. Index $q$ of the mode can be in principle found from the optimal
coupling  angle of incidence of the pump beam \cite{96a1GoIlSa}. However, this
possibility has not been realized in the present work. We just tried to excite and
analyze the modes with smaller $q$ index which are usually characterized by higher Q
and better coupling.

\section{\label{sect:spectra} Thermorefractive noise spectra}

\begin{figure}[htbp]
\scalebox{0.7}{\includegraphics{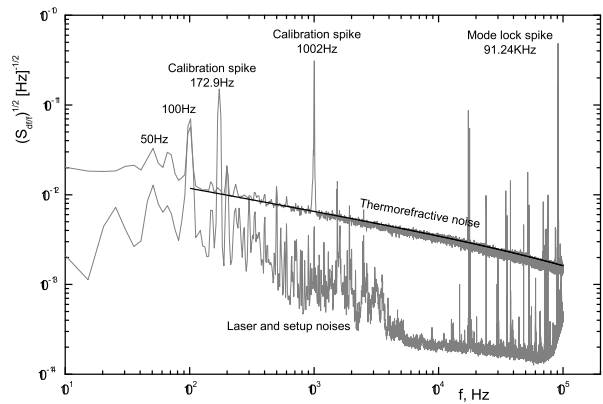}} \caption{\small Thermorefractive noise
in microsphere $(138\pm 8)\mu m$ in diameter, $l-m=4$. Lower curve represents setup
and laser noises, straight line ---  theoretical estimate.} \label{fig:lasernoise}
\end{figure}

Figure \ref{fig:lasernoise} presents typical high resolution spectrum of measured
noise in a microsphere 138$\mu$m in diameter and the spectrum of technical noises in
large 894$\mu m$ sphere measured as described in \ref{sect:hindrances}. Dashed line
shows theoretical curve for the spectrum calculated from recognized value of
$\ell-m=4\pm1$. Good agreement of experimental data with theory for the frequencies
$>100Hz$ is clearly observable.
 It can also be seen that the level of
combined technical noises, which include frequency and power fluctuations of the 
laser and electrical noises in circuits is nearly
$20dB$ lower than the observed effect, except for isolated spikes. 
Therefore, technical noises do not prevent the experimental test of thermorefractive model.

Calibrating spikes at $1002Hz$ for the smaller sphere and at $172.93Hz$ for the larger
one can be seen on the graphics. At low frequencies, peaks of technical noise at AC
$50Hz$--harmonics dominate. Additional sharp spikes could be seen at frequencies in
the order of $20kHz$ which were identified as acoustical resonances of laser mirrors,
responding to acoustical noise produced by laboratory equipment. One of these
resonances at $92kHz$ was used in feedback loop for active mode frequency
stabilization (see \ref{sect:calibration}). These spikes, except the specially excited
one at $92kHz$, are relatively small and narrow ($\sim 20$Hz) and could be seen only
due to high resolution of our measurements ($\sim 5Hz$ in the range of
$10\div10^5Hz$). They do not affect the result especially if additional averaging at
high frequencies is applied.
\begin{figure}[htbp]
\scalebox{0.3}{\includegraphics{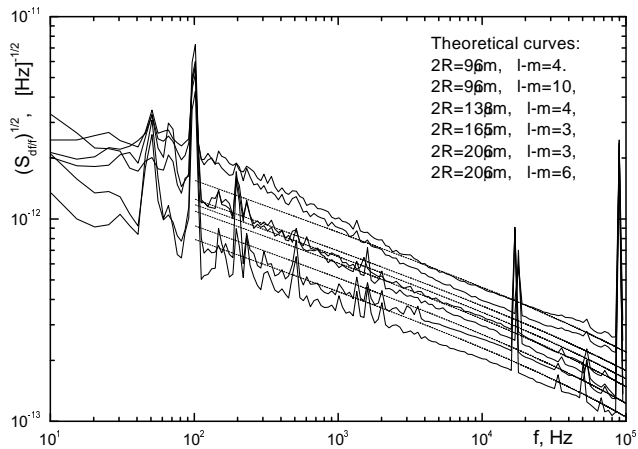}} \caption{\small Thermorefractive noise
in 4 microspheres for 6 different modes. Dashed lines --- theoretical curves obtained
for recognized modes' parameters. Larger noise corresponds to smaller sizes of
microspheres and smaller $\ell-m$ values} \label{fig:combined}
\end{figure}
Figure \ref{fig:combined} depicts results of measurement of calibrated spectra of
relative frequency fluctuations in four different microspheres for six different
modes. The sizes of microspheres and estimated mode parameters are given on the
figure.

Data for these graphics was obtained from the data presented on the previous figure
using uniform in log--scale averaging according to the following algorithm: if
$f_{min}(f_{max}/f_{min})^{j/N}<f_i<f_{min}(f_{max}/f_{min})^{(j+1)/N}$ then if
$n_j>0$, $f_j = \frac{1}{n_j}\sum f_i$; $S_j = \frac{1}{n_j}\sum S_i$, where $N$ is
the number of equidistant bins ($N=200$ for the presented figure) in log scale, $n_j$
is the number of points found in bin $j=0..N-1$, $f_{min}=10Hz$ and $f_{max}=10^5Hz$.
This approach allows one to estimate better the real spectra on higher frequencies
while preserving frequency resolution at low frequencies. To make the figure less
bulky, calibration peaks were digitally filtered out while processing.

\section{Discussion}
\label{sect:discussion}

We conclude that the results of measurements of
frequency noises in microspheres confirm the theoretical model of
thermorefractive fluctuations. This concerns both frequency dependence
and dependence on the mode parameters $\ell$ and $\ell-m$. The discrepancy
is less than 30\% in the frequency range of $3\times10^2\div10^5Hz$
($\Omega>1/\tau_R$), where the model and approximations used are
valid and which is free of installation noises. For frequencies
above $1kHz$, differences are within the limits of calibration and
WGM identification errors. Such a close agreement was obtained in
many resonators of average size of $120\div250\mu m$ in diameter.
In very small resonators at frequencies lower than $1kHz$, the
noise is up to two times higher, and the dependence on frequency
is more pronounced. In large resonators at lower frequencies the noise seems
to be smaller. However, as it was noted, the theoretical model
described below may be not valid when the frequency of noise is
comparable with the inverted time of temperature relaxation of the
whole microsphere via the microsphere stem ($R_s\sim 20\mu$ m).
The influence of the stem for smaller spheres due to the ratio $R_s/R$
is larger. This looks compatible to the observed systematical deviation
at low frequencies of experimental curves from the theoretical ones
obtained for thermally non-isolated spheres. Other mechanisms
of thermal noise, such as thermoelastic \cite{99a1BrGoVy} noise and thermal
fluctuations of effective eccentricity of microspheres leading to additional
frequency fluctuations proportional to $\ell-m$ can also contribute to
the effect. Additionally, there are other mechanisms of influence,
which can modify spectra, such as convective thermal relaxation
and influence of WGM locking scheme. 

It is worth noting, that from the displayed spectra it is possible to
calculate the spectra of microscopic fluctuations of temperature
in the volume, occupied by the e.m. field of WGM. Using the known
relation
\begin{equation}\label{link}
\frac{\delta \omega}{\omega}=-\frac{1}{n}\frac{dn}{dT}\delta T,
\end{equation}
one can rescale the spectra with the factor of
$1/\left(\frac{1}{n}\frac{\partial n}{\partial T}\right)=10^5K$
for fused silica. For instance, the mode of the $138\mu m$
resonator has an effective volume in the order of $4,3\times 10^{-9}$cm$^3$, which is
4 orders less than in former measurements of thermally induced
phase noise in long fibers \cite{WanserM}.

\appendix
\section{Calculation of thermorefractive noise in microspheres}
\label{sect:estimation}

To simplify the analysis we perform calculations below only for fundamental
whispering--gallery mode $TE_{\ell\ell1}$, which has the smallest volume of
localization. The result in principle may be extended to account for $TE_{\ell m q}$
modes with $q,\ell-m\sim1$. The field distribution of this mode may be approximated as
follows:
\begin{eqnarray}
& &{\bf \vec E}(r,\theta,\phi)\simeq {\bf \vec
E_\theta}(r,\theta,\phi) \frac{n \ell^{1/4}}{\sqrt{n^2-1}
R^{3/2}\pi^{3/4}}
\nonumber\\
&\simeq& e^{-\ell\cos^2\theta/2 + i\ell\phi}
\left\{\begin{array}{rl}
{{\rm j}_\ell(knr)/{\rm j}_\ell(knR)} \mbox{ for } r\leq R \\
e^{-\gamma(r-R)} \mbox{ for } r>R
\end{array} \right., \nonumber\\
knR&\simeq&\ell+1/2+1.8558(\ell+1/2)^{1/3}-\frac{n}{\sqrt{n^2-1}}.
\end{eqnarray}
However, even this approximation is too complex for analytical evaluation, so below we
shall use the following Gaussian approximation of radial dependence:
\begin{eqnarray}
& &{\bf \hat e}_\theta(r,\theta,\phi)\simeq \frac{1}{\pi
\sqrt{2bdR_0}}
e^{-\frac{(r-R_0)^2}{2b^2}-\frac{r^2\cos^2\theta}{2d^2} + i\ell\phi}\nonumber\\
& &knR_0\simeq\ell+1/2+0.71(\ell+1/2)^{1/3}\nonumber\\
& &d\simeq R_0\ell^{-1/2}\nonumber\\
& &b\simeq \frac{1}{kn}\sqrt{\frac{{\rm j}_\ell(knR_0)}{{\rm
j}^{\prime\prime}_\ell(knR_0)}}\simeq
0.84R_0\ell^{-2/3}\label{gaussian}.
\end{eqnarray}
This approximation describes rather adequately the distribution of
optical energy inside the resonator and allows one to calculate
Fourier integrals. Moreover, as it will be found below, because of
the small depth of the field (parameter $b$), radial distribution
practically does not influence the frequency fluctuations at
frequencies of interest.
\begin{eqnarray}
&&G(\vec\beta)= \frac{1}{2\pi^{2}
bdR_0}\int\limits_0^R\!\int\limits_0^{2\pi}\!\int\limits_0^\pi\!
e^{-\frac{r^2\cos^2\theta}{d^2}-\frac{(r-R_0)^2}{b^2}}
\nonumber\\
&&\times e^{-i\beta r(\cos\theta \cos\vartheta +
\sin\theta\sin\vartheta\cos(\phi-\varphi))} \sin\theta
d\theta\,d\phi\,r^2 dr  \nonumber\\
&\simeq& e^{-\frac{(\beta d\cos\vartheta)^2}{4}} e^{-\frac{(\beta
b\sin\theta)^2}{4}} \sqrt{\frac{1}{\pi\beta R_0\sin\vartheta}}.
\label{27}
\end{eqnarray}
To obtain this result, we used several approximations while
calculating the integrals. First integration was done over angle
$\theta$, taking  into account that the mode distribution is
narrow and hence $\sin(\theta)\simeq1$. Second integral over
$\phi$ leads to Bessel function of the first order, which was
approximated as spherical wave. Finally, integral over $r$ was
found considering only fast varying functions. Other physical
conditions were also used to neglect small terms: $(R-R_0)\ll R$,
$j(knR)\ll j(knR_0)$ and $\cos(\beta R)\simeq1$ --- due to
boundary conditions for thermal waves. In case of $\beta\to0$, the
last term in (\ref{27}) should be equal to 1 and the final
expression is incorrect due to used approximations for $\rm J_0$.
However, this case is not very interesting for us. To correct the
situation and obtain better approximation one may formally add 1
to the denominator.

Using (\ref{useful}) and keeping in mind additional factor 2 due to boundary
conditions we obtain:
\begin{eqnarray}
S_{\bar u}&=&\frac{8 k T^2 D}{\rho C}
\int\frac{\beta^2|G(\vec\beta)|^2}{a^4\beta^4+\Omega^2}\frac{d\vec\beta}{(2\pi)^3}\simeq \frac{2k T^2 D}{\rho C \pi^3R}\nonumber\\
&\times& \int\limits_0^\infty\!\int\limits_0^\pi e^{-\frac{(\beta
d\cos\vartheta)^2}{2}} e^{-\frac{(\beta
b\sin\theta)^2}{2}}\frac{\beta^3
d\vartheta\,d\beta}{D^2\beta^4+\Omega^2},
\end{eqnarray}
$\left[\int\limits_{0}^{\pi} e^{-x^2\cos^2(\vartheta)}
d\vartheta\simeq\frac{\sqrt{\pi}}{x}\right]$,
\begin{eqnarray}
S_{\delta\omega/\omega}(\Omega)&\simeq& \frac {\kappa T^2 D
}{\pi^{5/2} n^2 \rho C
R}\frac{2}{\sqrt{d^2-b^2}}\left(\frac{d\,n}{d\,T}\right)^2\nonumber\\
&\times&\int\limits_{0}^\infty \frac{\beta^2
e^{-\beta^2b^2/2}}{D^2\beta^4+\Omega^2}\frac{d\beta}{2\pi},
\label{ints}
\end{eqnarray}
\begin{eqnarray}
\int\limits^{\infty}_{0}\frac{\beta^2e^{-\beta^2b^2/2}}{D^2\beta^4+\Omega^2}\frac{d\beta}{2\pi}\simeq
\frac{\sqrt{2}}{4D^{3/2}\sqrt{\Omega}}\frac{1}{(1+(\Omega\tau_b)^{3/4})^2}.
\end{eqnarray}
This integral can be expressed through Lommel special functions, however, an
approximation is used here, which works well for $\Omega\tau_b<1$, where  $\tau_b
=(\pi/4)^{1/3}b^2/D$. Indeed, neglecting first item in denominator we obtain
high--frequency approximation with dependence $~\Omega^{-2}$ and neglecting
exponential approximation for lower frequencies $~\Omega^{-1/2}$ -- dependence is
obtained.

Finally:
\begin{eqnarray} S_{\delta\omega/\omega}(\Omega)&=& \left(\frac{d\,n}{d\,T}\right)^2\frac {\kappa
T^2\sqrt{\ell}}{\pi^{3/2} n^2R^2\sqrt{\lambda^*\rho
C\Omega}}\nonumber\\
&\times&\frac{1}{\sqrt{1-b^2/d^2}}\frac{1}{(1+(\Omega\tau_b)^{3/4})^2}.
\label{final}
\end{eqnarray}

Numerical analysis which is omitted here, shows, that for the
modes with $l\neq m$, the power spectral density of the
fluctuations is proportional to $\sqrt{2(l-m)+1}$. This dependence
is also confirmed by the fact, that azimuthal ``width'' of the
modes, and hence their effective volume, have the same dependence.

To account for finite size of microspheres, calculations using
(\ref{useful2}) should be performed. As at room temperature the
power radiated from the surface (Stephan--Boltzmann law) is much
lower than the heat exchange due to thermal conductivity (Fourier
law) the simplified boundary condition is used below:
\begin{equation}
\label{boundaryTC}
 \left. {\partial u(r,\theta,\phi,t)\over \partial r}\right|_{r=R}=0,
\end{equation}
\begin{eqnarray}
\Phi_{L,M,N}=C_{L,M,N}{\rm j}_L(\beta_{LN} r){\rm
P}^M_L(\cos\theta) \left\{^{\cos(M\phi)}_{\sin (M\phi)}\right.,
\end{eqnarray}
\begin{eqnarray}
C^2_{L,M,N}&=&\frac{2L+1}{\pi(1+\delta_{0M})}\frac{(L-M)!}{(L+M)!}\nonumber\\
&\times&\frac{\xi^2_{L,N}}{R^3(\xi^2_{LN}-L(L+1)){\rm
j}^2_L(\xi_{LN})},
\end{eqnarray}
\begin{eqnarray}
&&G_{L,N}=\frac{C_{L,0,N}}{\pi bdR_0}\\
&\times& \int\!\int
e^{-\frac{(r-R_0)^2}{b^2}-\frac{r^2\cos^2\theta}{d^2}} {\rm
j}_L(\beta_{LN} r){\rm P}_L(\cos\theta) r^2 dr \sin\theta
d\theta.\nonumber
\end{eqnarray}
Although only functions with $M=0$ lead to nonzero integrals, it is appropriate to
note, that functions with $M=2\ell$ could also be taken into account --- they lead to
the coupling between counter-propagating modes in the sphere and mode splitting.

To estimate this integral for small values of $\cos\theta\simeq\psi=\theta-\pi/2$ near
equator of microsphere, where the e.m. field is concentrated, the following
approximation for the Legendre polynomial may be used for \\ $L>0$
($P_0(\cos\theta)=1$):
\begin{eqnarray}
{\rm P}_{L}(\cos\theta)&\simeq& \sqrt{\frac{2}{\pi
L}}\left(1-\frac{1}{4L}\right)\nonumber\\ &\times&
\cos\left((L+\frac{1}{2})\psi + \frac{L\pi}{2}\right),
\end{eqnarray}
and for the spherical Bessel functions:
\begin{eqnarray}
\nonumber\\ j_L(z)&=&\frac{1}{z}\sin(z-L\pi/2), \nonumber\\
\xi_{LN}=\beta_{LN}R&\simeq& \frac{\pi(2N+L-1)}{2}, \nonumber\\
j_L(\xi_{LN})&\simeq&\frac{(-1)^{N-1}}{\xi_{LN}} \quad L=2K.
\end{eqnarray}
It is the crudeness of approximation for the roots of derivative of spherical Bessel
functions that limits the applicability of the final sum we obtain below. In this way,
the calculations below may be considered only as illustration.

\begin{eqnarray}
|G_{L,N}|^2&\simeq& \frac{2 {\cal R}(L,N)}{\pi^2 R^3}
e^{-\frac{(L+1/2)^2d^2}{2R_0^2}-\frac{\xi_{LN}^2b^2}{2R_0^2}}; \nonumber\\
{\cal R}(L,N)&=&\frac{(1-1/4L)^2(1+1/2L)}{1-L(L+1)/\xi^2_{LN}}
\quad \mbox{if} \quad L>0  \nonumber\\
{\cal R}(0,N)&=&\pi/4
\end{eqnarray}
In order to calculate these coefficients the approximation $\beta_{LN}(R-R_0)\ll 1$
was used. And finally:
\begin{eqnarray}
S_{\bar u}(\Omega)&\simeq& \frac{8 k T^2}{\pi^2 \rho D C R}\\
&\times& \sum_{K=0}^\infty\sum_{N=1}^\infty\frac{\xi_{LN}^2
e^{-\frac{(L+1/2)^2d^2}{2R_0^2}-\frac{\xi_{LN}^2b^2}{2R_0^2}}
}{\xi_{LN}^4+\frac{\Omega^2R^4}{D^2}} {\cal R}(L,N).\nonumber
\label{sums}
\end{eqnarray}

To calculate the sums in high--frequency approximation, we can
consider ${\cal R}(L,N)=1$. By making the following substitutions:
$x=(K+N)/\sqrt{2}$, $y=(K-N)/\sqrt{2}$ and $\tau_R=R^2/D$, we use
integrals instead of sums:
\begin{eqnarray}
S_u&\simeq& \frac {8\kappa T^2}{\pi^2 \rho C D R}\nonumber\\
&\times&\int\limits_{0}^{\infty}\int\limits_{-x}^{x}\frac{2\pi^2
x^2}{4\pi^4
x^4+\Omega^2\tau^2_R} e^{-\pi^2 x^2 b^2/R^2}e^{-(x+y)^2/\ell} dy dx\nonumber\\
&\simeq& \frac {8\kappa T^2}{\pi^2\rho C D R}\sqrt{\frac{\ell}{8\pi}}\nonumber\\
&\times& \int\limits_{0}^{\infty}\frac{t^2}{t^4+\Omega^2\tau_R^2}
e^{-t^2 b^2/(2R^2)}{\rm
erf}\left(\frac{\sqrt{2}t}{\pi\sqrt{\ell}}\right)dt.
\end{eqnarray}
This integral coincides with (\ref{ints}) obtained for infinite media, if
erf--function, which is practically equal to unity for $t>1/\sqrt{\ell}$,
is ignored. However, spectral densities described by integral (\ref{ints})
and sum (\ref{sums}) are different at frequencies for $\Omega<1/\tau_R$
where (\ref{sums}) is finite for $\Omega\to 0$ (the term with two sums
is close to unity).

\section{Relation between regular calibration signal and noise spectrum}

To find the link between the amplitude of harmonic frequency
change, corresponding peak in calculated spectral density and
noise spectral density, we consider the Fourier transform for
regular and chaotic processes.  Let $x(t)$ denotes chaotic process
with correlation function of
\begin{equation}
B(\tau)=<x(t)x(t+\tau)>=\int\limits_{-\infty}^{\infty}S^{\pm}(\Omega)e^{i\Omega\tau}\frac{d\Omega}{2\pi},
\end{equation}
where $S^{\pm}(\Omega)$ stands for ``double--sided'' spectral
density which is symmetrical and two times less for positive
frequencies than ``one--sided'' $S(\Omega)$. Let $T$ denotes the
length of the ``window'' $W(t)$ in time domain, that is the
duration of a set $\{x(t)\}$ (or $\{x_i\}$ in discrete case) for
which the Fourier transform is carried out. Now we shall consider
the transform with the square window  $w(t)=1$ for
$t\in[-T/2,T/2]$. Fourier transform for $x(t)$ is:
\begin{equation}
X_j=\int\limits_{-T/2}^{T/2}W(t)x(t)e^{-i\Omega_j t}dt,\;\;\;\;\; \mbox{where}\;\;\;
\Omega_j=\frac{2\pi j}{T}.
\end{equation}
In the procedure of spectrum estimation the spectra $X_j(\Omega)$
are calculated for small intervals of data and then averaged,
which in our notations corresponds to the calculation of the
quantity $\langle|X_j|^2\rangle$.
\begin{eqnarray}\label{programresult}
\langle|X_j|^2\rangle&=&\int\limits_{-T/2}^{T/2}\langle x(t)x(t')\rangle e^{i\omega_j(t-t')}dtdt' \nonumber\\
&=&\int\limits_{-\infty}^{\infty}
S(\omega)\frac{4\sin^2[(\omega_j-\omega)\frac{T}{2}]
}{(\omega_j-\omega)^2} \frac{d\omega}{2\pi}.\\ \nonumber
\end{eqnarray}
Calibrating peak in our case corresponds to frequency
$\omega_j>>T^{-1}$. Then, the core in the latter integral has
sharp maximum at $\omega=\omega_j$ and
$\langle|X_j|^2\rangle=TS(\omega_j)$.

Let now $y(t)=Y_0\cos(\Omega t)$ represents a regular harmonic
signal. The Fourier transform for such signal for square window
will look as follows
\begin{eqnarray}
Y_s&=&\int\limits_{-T/2}^{T/2}Y_0\cos (\Omega t)e^{-i\Omega_jt}dt,
\end{eqnarray}
leading to $Y_{sj}=Y_0 T/2$. Now we compare the results of
spectrum estimation procedure (\ref{programresult}) and the
results of Fourier transform for harmonic signal:
\begin{displaymath}
Y_{sj}^2=\frac{Y_0^2T^2}{4} \Leftrightarrow
TS^{\pm}(\omega_j)=\langle|X_j|_s^2\rangle.
\end{displaymath}
Here one can easily see the link between the amplitude of harmonic signal and
corresponding peak in spectral density. If the harmonic signal has the dimension of
frequency and the desired spectral density represents relative frequency changes, then
the final relation, corresponding to one--sided spectral density of calibration spike
with square window in Fourier transform, will be
\begin{equation}
\sqrt{S_{\delta
f/f}}\Leftrightarrow\frac{Y_0}{f_0}\sqrt{\frac{T}{2}}
\left[\frac{\mbox{1}}{\sqrt{Hz}}\right],
\end{equation}
where $f_0$ --- is optical frequency. The same calculations for the Hann window give:
\begin{equation}
\sqrt{S_{\delta
f/f}}\Leftrightarrow\frac{Y_0}{f_0}\sqrt{\frac{T}{3}}
\left[\frac{\mbox{1}}{\sqrt{Hz}}\right].
\end{equation}

\section{Formulas for WGMs identification}

To identify the parameters of whispering--gallery modes from photographs the following
approximation for half--width of WGM belt was used:
\begin{equation}
  \Delta\theta = \sqrt{2(\ell+1/2-m)/\ell}.
\end{equation}
To calculate  $\ell$--index from the radius of microsphere we
found by numerical approximation the following approximated
formulas valid for $R\gg\lambda$:
\begin{eqnarray}
q=1: &l=-0.5+t_{lq}-2.287t_{lq}^{1/3}+0.1718t_{lq}^{-1/3},\\
\nonumber q=2:
&l=-0.5+t_{lq}-4.617t_{lq}^{1/3}+0.6944t_{lq}^{-1/3},\\ \nonumber
q=3: &l=-0.5+t_{lq}-6.895t_{lq}^{1/3}+1.518t_{lq}^{-1/3},\\
\nonumber q=4:
&l=-0.5+t_{lq}-9.190t_{lq}^{1/3}+2.632t_{lq}^{-1/3},
\end{eqnarray}
where
\begin{eqnarray}
t_{lq}=\frac{2\pi n_i
R}{\lambda_{lmq}}+\frac{pn_i}{\sqrt{n_i^2-1}}.
\end{eqnarray}
These formulas allow to get an estimate of $\ell$ and $\ell-m$
values from the radius of the microsphere and width of the belt of
the mode visualized due to residual surface scattering.
\bigskip\bigskip\bigskip\bigskip

\acknowledgments     
We are grateful to professor V.B.Braginsky, whose initial idea to measure
thermorefractive noise in microspheres has embodied in this paper. Special thanks to
professors S.P.Vyatchanin and F.Ya. Khalili for stimulating and fruitful discussions.
This work was supported partially by NSF grant \#PHY-0098715 and Russian Foundation
for Basic Research grant \#00-15-96706.


\bibliography{thermic}

\end{document}